\documentclass{article}

\usepackage{arxiv}

\usepackage[utf8]{inputenc} 
\usepackage[T1]{fontenc}    
\usepackage{hyperref}       
\usepackage{url}            
\usepackage{booktabs}       
\usepackage{amsfonts}       
\usepackage{nicefrac}       
\usepackage{amsmath}
\usepackage{microtype}      
\usepackage{cleveref}       
\usepackage{lipsum}         
\usepackage{graphicx}
\usepackage{natbib}
\usepackage{doi}

\usepackage{hyperref}

\usepackage{subcaption}
\usepackage[T1]{fontenc}

\usepackage{booktabs}
\usepackage{siunitx}

\title{Outlier-Based Domain of Applicability
Identification for Materials Property
Prediction Models}

\author{ {
\hspace{1mm}Gihan Panapitiya}\\
	Pacific Northwest National Laboratory\\
	Richland, WA 99354 \\
	\texttt{gihan.panapitiya@pnnl.gov} \\
	\And
	{
 \hspace{1mm}Emily Saldanha} \\
	Pacific Northwest National Laboratory\\
	Richland, WA 99354 \\
	\texttt{emily.saldanha@pnnl.gov} \\
}

\renewcommand{\shorttitle}{Domain of Applicability}

\hypersetup{
pdftitle={A template for the arxiv style},
pdfsubject={q-bio.NC, q-bio.QM},
pdfauthor={David S.~Hippocampus, Elias D.~Striatum},
pdfkeywords={First keyword, Second keyword, More},
}

\begin{document}
\maketitle

\begin{abstract}
Machine learning models have been widely applied for material property prediction. However, practical application of these models can be hindered by a lack of information about how well they will perform on previously unseen types of materials. Because machine learning model predictions depend on the quality of the available training data, different domains of the material feature space are predicted with different accuracy levels by such models. The ability to identify such domains enables the ability to find the confidence level of each prediction, to determine when and how the model should be employed depending on the prediction accuracy requirements of different tasks, and to improve the model for domains with high errors. In this work, we propose a method to find domains of applicability using a large feature space and also introduce analysis techniques to gain more insight into the detected domains and subdomains.
\end{abstract}

\keywords{Domain of Applicability \and Machine Learning}

\section{Introduction}

The use of data-driven approaches for material property prediction and material design have become increasingly common. Because of the dependency between the quality and availability of training data and the resulting model predictions, trained property prediction models will achieve different levels of accuracy on different regions of the material feature space. The ability to identify such domains helps with understanding the limitations of a given model. Such information is needed to ensure that the model is applied only where valid predictions can be obtained and that unwarranted trust is not placed in predictions for which the model cannot function well. In this work, we use the prediction of molecular solubility  as a test case to develop a domain of applicability (DoA) detection method and provide analysis techniques to gain further insight into the identified domains.

Solubility prediction plays a vital role in many disciplines including drug discovery~\cite{Christel_2018, Lipinski_2001,Li_2012}, medicine~\cite{savjani2012, Cisneros2017}, fertilizers~\cite{Guo2020}, and energy storage systems~\cite{li_2015, Yan_2021}. Despite efforts over many decades, there are still limitations in the development of highly accurate predictive models that can function well across a diverse range of molecules. However, as is the case with any predictive task, there are certain regions of molecular structure space where the predictions do achieve high accuracy. The ability to identify the regions in the molecular descriptor space that have sufficient predictive accuracy has two main benefits: 

\begin{enumerate}
    \item \textbf{The reliability of predictions for artificially designed molecules can be assessed.} Machine learning-based molecular design and discovery pipelines rely on the application of predictive models to novel molecular structures. Without knowing the reliability of a model associated with a particular prediction, we are not in a position to make recommendations regarding the corresponding molecule - for example suggesting to proceed with synthesizing the molecule. DoA analysis provides the ability to assess the likelihood of erroneous predictions on novel molecular structures. 
    \item \textbf{Improvements to models can be targeted by understanding their current limitations.} By analyzing the regions of molecular feature space that are not well predicted by the model, the model developers can target additional data collection or modeling approaches to address these weaknesses.
\end{enumerate}

In this work we propose a pipeline for identifying domains in molecular feature space associated with different levels of predictive accuracy. The major contributions of our work include: (1) a method to find domains of applicability which we show to be effective for data not previously seen by the model, (2) an outlier-inlier separation method based on an \textit{outlierness} score, (3) clustering-based analysis methods to gain more insight into the detected domains, 4) insights into the relationships between applicability domains and predictive errors of the model.

We start with first identifying molecules that behave like outliers in terms of the model predictive accuracy. Such molecules are likely to identify the boundary regions of the model's domain of applicability. We define these DoA boundaries by comparing the distributions of molecular descriptors corresponding to outlier and non-outlier molecules to identify feature values corresponding to higher outlier likelihood. We show the validity of the identified domains by demonstrating that model accuracy improves as out-of-domain molecules are removed from the evaluation set. Finally, we identify a set of molecules which behave like outliers in terms of predictive accuracy but are not outside the domain of applicability in terms of their structural features. We perform analysis to understand the factors that lead to the anomalous predictive behavior for these molecules.

The term Domain of Applicability has been used in slightly different contexts in the literature. In one context, the applicability domain is identified as the range of feature space of the materials used to train a predictive model \cite{Jaworska2005, Sahigara2012}. According to this definition, a query material is identified as within the domain of applicability if it is structurally similar to those used in the training. Some authors only use independent physico-chemical properties for this comparison while some also include the prediction or independent target property \cite{Jaworska2005}. Various methods based on feature ranges, distance between data points, and probability density distribution have been used to find the domains of applicability~\cite{Jaworska2005, Sahigara2012, Sheridan2004, Tetko_2008}. In range based methods, a hyper-rectangle defining the domain is constructed using either raw features or transformed features such as principle components \cite{Jaworska2005}. In the distance-based methods, the distance from the test data point to a point representing the reference data is measured. The centroid of the reference data, distance to K-nearest neighbors in the reference data, and average distance between the test point and the data points in the reference data are among the approaches used to select the reference point \cite{Jaworska2005, Sahigara2012}. The test points with distances less than a predefined threshold value are considered to be part of the domain of the reference data.

Rather than simply identifying regions near the training data, our approach aims to identify domains in the feature space where models can achieve high prediction accuracy. Prior efforts have approached DoA identification in a similar context. For example, \citet{sutton_identifying_2020} presented a method based on subgroup discovery~\cite{Boley2012, van2012, nguyenV2015} to find such domains with high predictive accuracies based on an impact metric. In our work, we use a different strategy to identify the subgroups and rank order them in terms of their impact on the prediction accuracy. Our domain identification is based on the differences in the feature distributions of outlier and inlier molecules. Similarly to \citet{sutton_identifying_2020}, we define domains in the form  of individual feature inequality constraints (i.e nAtom < 200 indicates that the molecules with  less than 200 atoms are in-domain while larger molecules are out-of-domain). However, in our work, these feature  inequalities are rank ordered using an \textit{outlier-ness} score of the molecules that do not satisfy the inequality. This rank ordering ensures that we can identify the features that can most effectively separate the regions of high model accuracy. We have released our code at https://github.com/pnnl/doa.


\section{Data}
Our dataset is composed of three prior datasets consisting of experimental aqueous solubility measurements: the SOMAS dataset (10,162 molecules)~\cite{gao_somas_2022, Panapitiya2022}, AqSolDB (4425 molecules) \cite{sorkun_2019} and the Cui dataset (7283 molecules) \cite{cui2020}. We first merge the three datasets and resolve duplicate molecules using a method described in the Supporting Information. Next, we  divide the full dataset into train and test sets, where the train set is used to determine the rules to identify the domain of applicability and the test set is used to evaluate validity of the identified domains. The train and test sets consist of 20,229 and 1641 molecules respectively. To avoid confusion with another train set used in our method, we refer to this train set as the \textit{full train} set. For training machine learning models based on this data, we calculate 721 derived molecular descriptors using the Mordred package~\footnote{https://github.com/mordred-descriptor/mordred}\cite{Moriwaki_2018} and 87 3D and fragment-based descriptors using our own package \cite{Panapitiya2022, Panapitiya2018}.

\section{Methods}\label{sec:methods}
\subsection{\textbf{Outlier Detection}}\label{sec:ol-detect}
Our approach starts by first identifying the outlier molecules in the dataset where we consider outliers to be molecules for which the model can not predict well or consistently. 

\begin{figure*}[!t]
     \centering
    \begin{subfigure}[b]{0.31\textwidth}
         \centering
         \includegraphics[width=\textwidth]{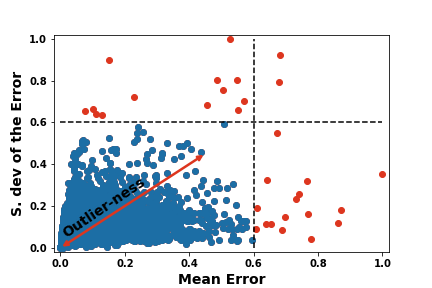}
         \caption{}
         \label{fig:mean-sdev}
     \end{subfigure}
         \begin{subfigure}[b]{0.31\textwidth}
         \centering
         \includegraphics[width=\textwidth]{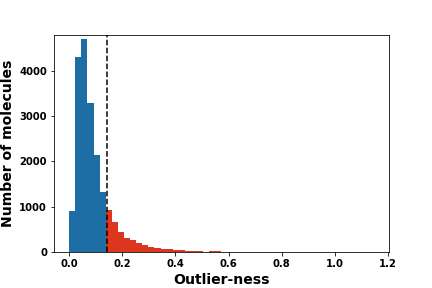}
         \caption{}
         \label{fig:dist-olness}
     \end{subfigure}
     \begin{subfigure}[b]{0.31\textwidth}
         \centering
         \includegraphics[width=\textwidth]{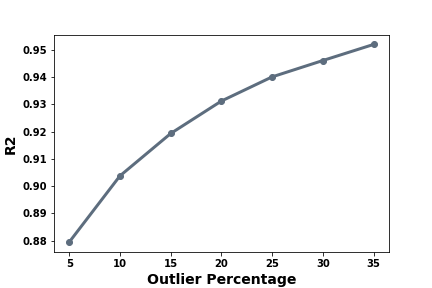}
         \caption{}
         \label{fig:olp-vs-r2}
     \end{subfigure}

    \caption{(a) Mean error versus the standard deviation of the error for each molecule in the entire dataset. (b) Distribution of \textit{outlier-ness} for all molecules. (c) The achieved cross-validated $R^2$ values as a function of the percent of molecules removed based on \textit{outlier-ness}.}
        \label{fig:ol-detect}
\end{figure*}

The outlier detection method is based on an algorithm described in  \citet{Cao2017}. We perform multiple repetitions of training a machine learning model to predict solubility from the molecular features by splitting the \textit{full train} set into random train and validation sets multiple times. For each trained model, the resulting predictive errors measured using root mean squared error (RMSE) on the molecules in the validation set are collected. Consequently, multiple error results exist for each molecule as they each are sampled into the validation set multiple times. We define outlier molecules as those that are associated with large mean error, large variability in the error, or both. The machine learning algorithm we used as the prediction model is ExtraTreesRegressor as implemented in the \textit{scikit-learn} package~\footnote{https://scikit-learn.org/stable/}. Training was performed using the default parameters, each time taking a different random train and validation split. We perform 1000 such repetitions and find that each molecule has been predicted as part of the validation set on average 100 times with a standard deviation of about 10 times. The minimum number of predictions made for any molecule is 59 which provides sufficient signal to quantify the prediction error and its variability. The mean and the standard deviation of the prediction errors of all the molecules in the train set are shown in Figure \ref{fig:mean-sdev}.

The next task is to define a criteria to separate outliers and inliers. The method given in the original approach~\cite{Cao2017} involves dividing the  mean error - standard deviation of the error space into four quadrants as shown in Figure~\ref{fig:mean-sdev} and labeling molecules belonging to farthest quadrants from the origin (colored in red in Figure \ref{fig:mean-sdev}) as outliers. In contrast, we seek to separate outliers by defining a single \textit{outlier-ness} metric which takes into account both the mean error and the error variability. The outlier-ness is found by first scaling the mean errors and error standard deviation so that these values are between 0 and 1 using MinMax scaling (Figure \ref{fig:mean-sdev}). \textit{Outlier-ness} is then calculated as the Euclidean distance from the origin to each data-point in the mean-variation space as shown by the red arrow in Figure \ref{fig:mean-sdev}. The distribution of the resulting \textit{outlier-ness} values for all the molecules are shown in Figure \ref{fig:dist-olness}. 

Next we aim to define a threshold for \textit{outlier-ness} such that a certain proportion of molecules is selected as outliers (for example the red shaded region in Figure \ref{fig:dist-olness}). In Figure~\ref{fig:olp-vs-r2}, we plot the achieved $R^2$ of the inlier data as a function of what proportion of the data is removed as outliers. This approach allows us to associate the chosen inliers with a particular prediction accuracy. For this study, we selected 20\% as the outlier percentage which results in $> 0.92$ $R^2$ value for the corresponding inliers. 


\subsection{\textbf{Finding Domains of Applicability}}\label{sec:doa-detect}
Once we identify the outliers and inliers within the dataset, we are ready to identify the domains of applicability for the model. Rather than splitting the data into single set of out-of-domain (OOD) and in-domain (ID) molecules, we aim to find increasingly strict ID regions of molecular space for which we can expect to achieve increasing levels of model accuracy. These domain regions are defined in terms of threshold values for different molecular descriptors such that the cumulative application of additional thresholds leads to the smaller and smaller hyper-rectangular regions of molecular feature space.

To define the boundaries of these regions, we aim to identify thresholds beyond which molecules have a high likelihood of being outliers. To identify the relevant molecular descriptors, we first address the issue of high levels of correlation among the descriptors. If two descriptors are correlated with a Pearson coefficient greater than 0.95, we randomly remove one of them from the dataset to ensure that we identify the unique factors driving the model behavior. As shown in Table 
\ref{table:doa-table}, our goal is to find a list of descriptors along with threshold values that separate ID and OOD molecules for each domain. Details regarding the definitions of the descriptors can be found in the Supporting Information, Mordred documentation \cite{Moriwaki_2018} and Reference \cite{Panapitiya2022}. In what follows, we describe the method of building this table.

First, we aim to select which descriptors distinguish the OOD molecules from the ID molecules. As an initial screening step, we identify descriptors for which the outlier molecules have a larger range of values than the inliers on either the high or the low end. The molecular descriptors are first scaled to zero mean and unit variance so that the differences are comparable across descriptors and are not driven by the natural magnitude of the descriptor values. 
Next, for each descriptor $X$, we calculate the difference between maximum and minimum values of outlier values ($X_{out} = \{X_i \vert i \in \mathrm{outliers} \}$) and inlier values ($X_{in} = \{X_i \vert i \in \mathrm{inliers} \}$). 

\begin{equation}
\begin{split}
        \Delta_{max} & = \max X_{out} - \max X_{in} \\ 
        \Delta_{min} & = \min X_{out} - \min X_{in} \\ 
        \end{split}
\end{equation}

It should be noted that we only consider cases which satisfy $\Delta_{max} > 0$ or $\Delta_{min} < 0$. That is, when either the maximum outlier value exceeds that of the inliers or the minimum outlier value lies beneath that of the inliers.

\begin{figure*}[!t]
     \centering
     \begin{subfigure}[b]{0.49\textwidth}
         \centering
         \includegraphics[width=\textwidth]{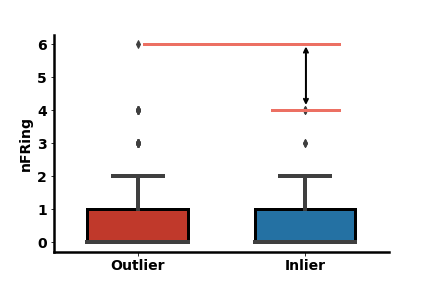}
         \caption{}
         \label{fig:th-difference}
     \end{subfigure}
     \begin{subfigure}[b]{0.49\textwidth}
         \centering
         \includegraphics[width=\textwidth]{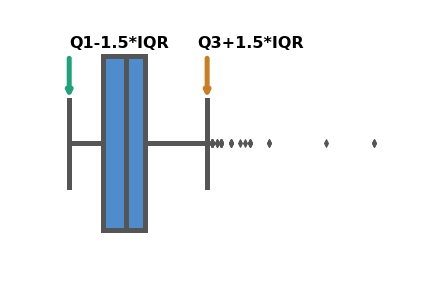}
         \caption{}
         \label{fig:th-ths}
     \end{subfigure}
     
    \caption{Threshold selection. (a) We use the difference between extreme values (maximum and minimum) as an initial filtering step to determine which descriptors may be relevant for defining the DoA boundaries. (b) For the selected descriptors, we use the whiskers of the inlier descriptor distribution as the thresholds that separate ID and OOD regions.}
        \label{fig:th-detect}
\end{figure*}

For each descriptor that meets this initial filtering criteria, we then seek to define a threshold that will be used to divide ID molecules from OOD molecules. We use the distribution of the inliers to select the threshold, specifically leveraging the first (Q1) and third (Q3) quartiles and the interquartile range (IQR) of the entire descriptor value distribution (that is, the descriptor values of inliers and outliers). We set a threshold which is calculated as either (Q1+1.5 $\times$ IQR) or (Q3-1.5 $\times$ IQR) depending on whether the outlier values exceed or lie beneath the inliers (Figure \ref{fig:th-ths}). The threshold value, $X_{t}$ , for each descriptor is selected using the following criteria,

\begin{equation}
\begin{split}
         X_{t} & = Q3+1.5 \times IQR, \text{if } |\Delta_{max}|  > |\Delta_{min}|\\
        X_{t} & = Q1 - 1.5 \times IQR, \text{if } |\Delta_{max}|  < |\Delta_{min}|\\
\end{split}
\end{equation}

Once we determine the threshold for each descriptor, the next task is to find the relative importance of each descriptor in separating ID and OOD molecules. We achieve this by calculating the mean \textit{outlier-ness} of OOD molecules defined for each descriptor. For example, for the descriptor \textit{Mpe}, descriptor values greater than 0.875 correspond to OOD molecules. We calculate the mean \textit{outlier-ness} of these OOD molecules and use it as a measure of importance of \textit{Mpe} relative to the other descriptors. The descriptors in Table \ref{table:doa-table} are arranged in the descending order of the mean \textit{outlier-ness} of the OOD molecules.
The direction specifies whether the property value of the molecule should be equal to, less than or equal to, or greater than or equal to the threshold in order to consider it part of the domain of applicability. For example, we can define the first domain of applicability as molecules for which \textit{NssssSn} == 0, while the second domain of applicability is molecules for which \textit{NssssSn} == 0 and \textit{Mpe} $>=$ 0.875.

\begin{table}[!thb]
\caption{ DoA table. Descriptor thresholds defining the domain of applicability and resulting test set performance ($R^2$ and RMSE) after limiting the data to this domain. }
\centering

\begin{tabular}{cllccrr}
\toprule
 Domain &       Descriptor &           Threshold & Direction &  Test OOD \% &       R\textsuperscript{2} &     RMSE \\
\midrule
      1 &    NssssSn &     0.00 &         == &      0.06 & 0.78 &  1.04 \\
      2 &        Mpe &     0.87 &         >= &      0.18 & 0.78 &  1.04 \\
      3 &        IC2 &     5.45 &         <= &      0.30 & 0.79 &  1.02 \\
      4 &  n10AHRing &     0.00 &         == &      0.37 & 0.79 &  1.02 \\
      5 &        IC3 &     6.05 &         <= &      0.55 & 0.79 &  1.02 \\
      6 &       bpol &    55.58 &         <= &      2.86 & 0.80 &  0.97 \\
      7 &    n5FRing &     0.00 &         == &      2.86 & 0.80 &  0.97 \\
      8 &     nFRing &     2.50 &         <= &      2.86 & 0.80 &  0.97 \\
      9 &     SsssNH &     0.00 &         >= &      2.86 & 0.80 &  0.97 \\
     10 & nG12aHRing &     0.00 &         == &      2.86 & 0.80 &  0.97 \\
     11 &       TIC5 &   420.58 &         <= &      4.20 & 0.80 &  0.96 \\
     12 &     ATSC0v &  4399.51 &         <= &      5.00 & 0.80 &  0.96 \\
     13 &     SsAsH2 &     0.00 &         == &      5.00 & 0.80 &  0.96 \\
     14 &         n5 &     0.00 &         == &      5.79 & 0.80 &  0.94 \\
     15 & nG12AHRing &     0.00 &         == &      5.91 & 0.80 &  0.94 \\
     16 &         n6 &     0.00 &         == &      5.97 & 0.80 &  0.95 \\
     17 &      ATS1m &  7851.87 &         <= &      6.52 & 0.80 &  0.94 \\
     18 &        fMF &     0.75 &         <= &      6.52 & 0.80 &  0.94 \\
     19 &      nRing &     6.00 &         <= &      6.70 & 0.80 &  0.94 \\
     20 &      ATS2m & 11514.37 &         <= &      7.56 & 0.80 &  0.94 \\
     21 &      ATS0p &   127.37 &         <= &     10.60 & 0.82 &  0.90 \\
\bottomrule
\end{tabular}

\label{table:doa-table}
\end{table}

\begin{figure*}[!t]
    \centering
         \includegraphics[width=.8\textwidth]{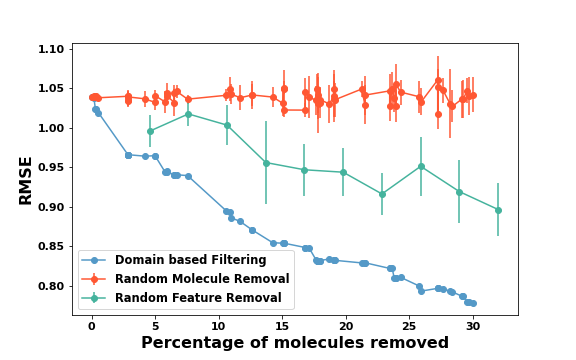}
    \caption{Reduction in the test set RMSE with the removal of data corresponding to different domains (blue) compared with the removal of random groups of molecules (red) and removal based on random selection of features (green). }
        \label{fig:fbf}
\end{figure*}

The application of these threshold filters should define increasingly small regions of molecular space for which the model is able to make increasingly accurate predictions. In order to evaluate the validity of the domains, we observed the RMSE of the test set predictions after removing OOD molecules corresponding to each domain from the test set. This shows the expected level of accuracy the model can achieve when we limit its application to just those molecules falling within the selected applicability domain. The blue curve in Figure \ref{fig:fbf} shows our results. Each point corresponds to a domain in Table \ref{table:doa-table}. We observe that as we narrow the domain (as we go down the DoA table), we achieve better prediction accuracy on the remaining ID molecules. We compare this result with the case in which we remove sets of random molecules of the same size from the test set and make predictions on the remaining data. We find that predictive performance is not affected by this random removal. Additionally, to validate the \textit{outlier-ness}-based ordering of the descriptor threshold application, we also tested the effect of randomly sampling descriptors to define the domains. The resulting test RMSE values shown by the green curve shows that the \textit{outlier-ness}-based domain ranking is crucial.

\subsection{Subdomains}\label{sec:subdomains}
\begin{figure*}[!t]
    \centering
         \includegraphics[width=1\textwidth]{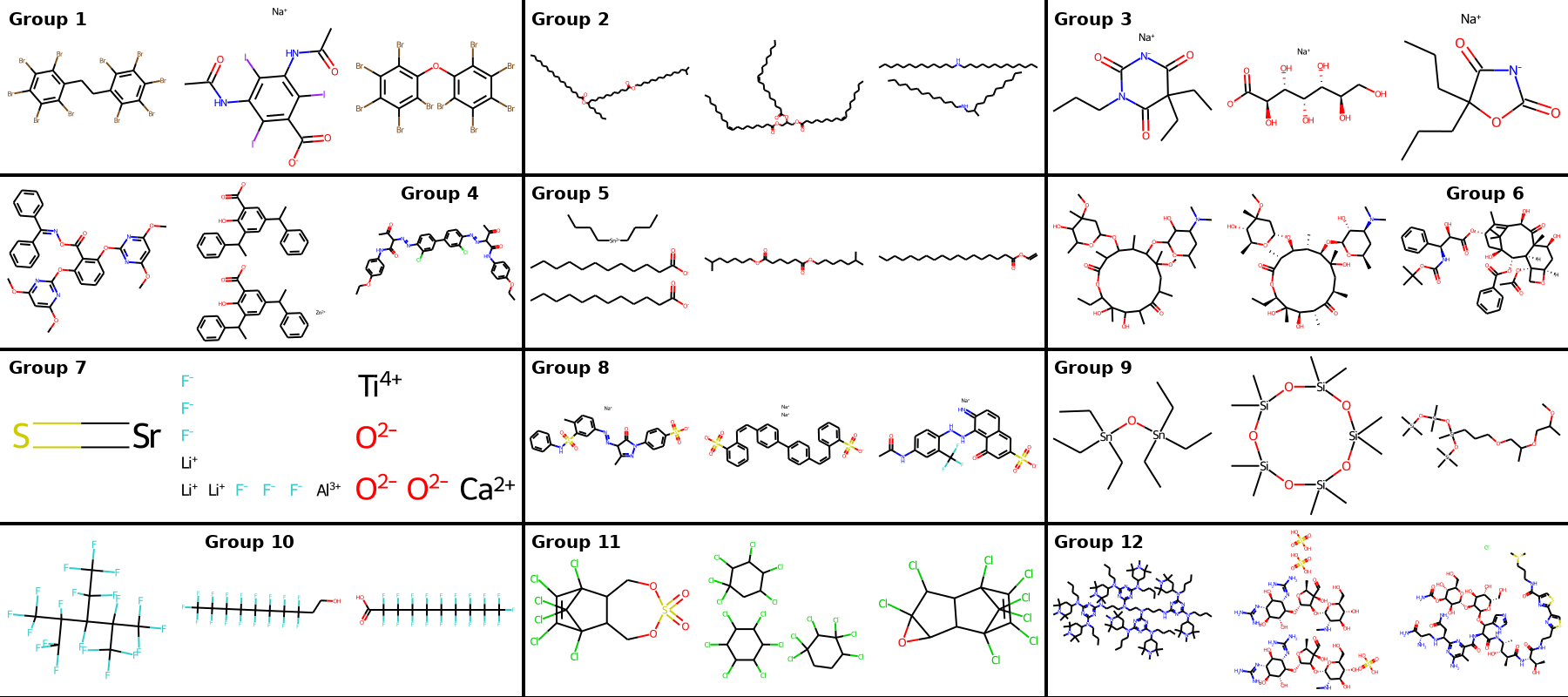}
    \caption{Groups of OOD molecules identified using clustering.}
        \label{fig:ood-groups}
\end{figure*}

We next seek to qualitatively understand the types of molecules which are determined to be in-domain and out-of-domain by this method.  While the descriptor thresholds provide information about the boundaries of the domains, they do not give us the full picture of the distributions of molecules within each group, especially for domains which leverage a small number of descriptors to define the domain boundaries. To gain more insight, we employ K-means clustering to identify subgroups inside the chosen domains. In order to reduce the effect of the random nature of the clustering method, we perform clustering several times and select the molecules that get grouped together most frequently. The details of this approach are explained further in Supporting Information. In Figure \ref{fig:ood-groups} we show examples of twelve molecular groups which are outside of Domain 21 in the DoA table (Table \ref{table:doa-table}) to provide a qualitative picture of the types of molecules which are out-of-domain for the model.

\begin{figure*}[!t]
    \centering
         \includegraphics[width=1\textwidth]{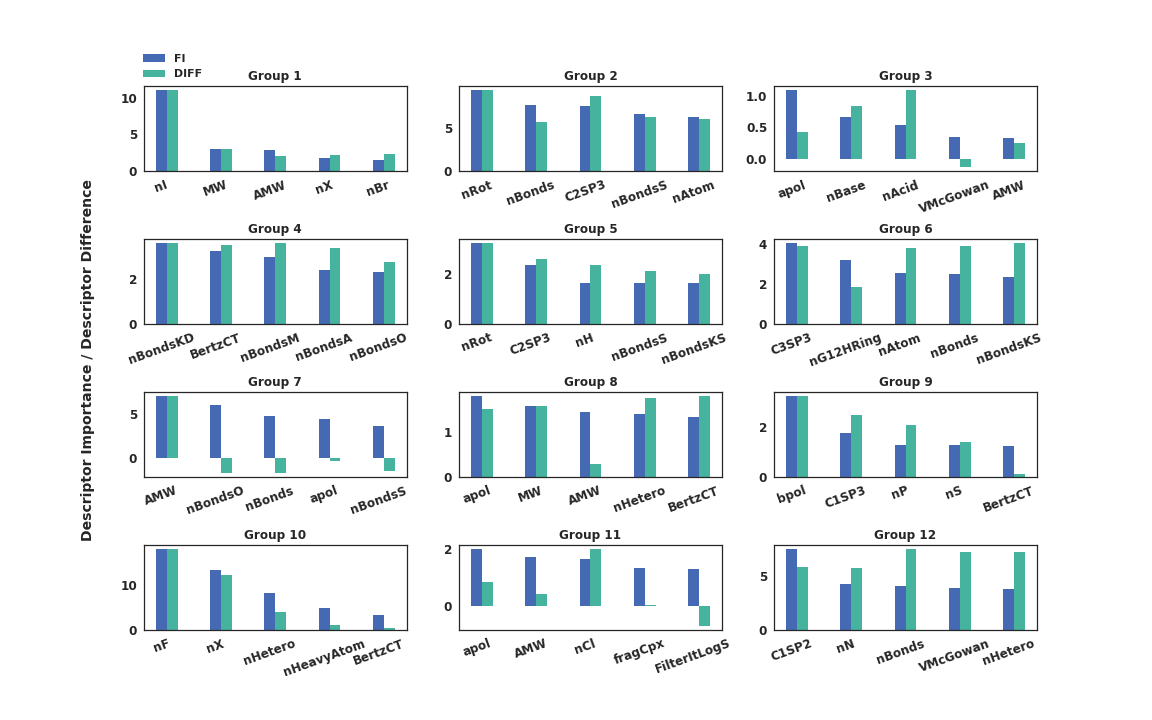}
    \caption{The distinguishing characteristics of each cluster of OOD molecules. The blue bars indicate the relative descriptor importances for the ExtraTreesClassifier when classifying the molecule groups in Figure \ref{fig:ood-groups} versus the other molecules. The green bars indicate the difference between mean descriptor values of molecule groups in Figure \ref{fig:ood-groups} and the other molecules.}
    \label{fig:feature-difference-expl}
\end{figure*}

For some of the clusters, it is easy to recognize their distinguishing characteristics just by visual inspection. For example, Group 2 contains molecules with long carbon chains and Group 10 consists of molecules containing fluorinated carbon chains. However, for other groups, the distinguishing characteristics are less clear when only looking at a few examples. Therefore, we seek to quantitatively define the prominent descriptors that characterize each group. We leverage a classification method with the capability to rank descriptor importances to identify the molecular descriptors that distinguish the subdomain from the rest of the molecules. The classifier we used was the ExtraTreesClassifier as implemented in the \textit{scikit-learn} package with default parameters. The molecules belonging to a given group in Figure \ref{fig:ood-groups} are assigned as the positive class and all the other molecules in our \textit{full train} set (defined in Section \ref{sec:ol-detect}) are assigned as the negative class. While the ExtraTreesClassifier can identify which descriptors were most important in distinguishing the molecules, it cannot automatically identify how the in-group molecules differ from the others in terms of those features. Therefore, we also calculated the difference between the mean values of in-group and out-of-group descriptors ($\bar{X}_{in-group} - \bar{X}_{out-of-group}$) to identify whether the in-group molecules have higher or lower values of each feature than the typical molecule.  In Figure \ref{fig:feature-difference-expl} we show five most important descriptors for each group (blue bars) determined by the classifier along with the corresponding mean descriptor differences (green bars). To aid subdomain interpretability for the purposes of this analysis, we focus on easy-to-interpret descriptors rather than more complex and theoretical descriptors. These easy-to-interpret descriptors were chosen by inspection. Before finding the feature differences, we have scaled each descriptor so that it has a zero mean and a standard deviation of 1 (standard scaling), such that the descriptor differences can be interpreted as standard deviations. The descriptor importances are also scaled so that the largest value is equal to the largest descriptor difference value. This scaling does not affect the interpretation of descriptor importances, as these values can only be compared relatively within the same group. 



We will present several examples to demonstrate how we can use the information in Figure \ref{fig:feature-difference-expl} to gain more insight on each group. Such analysis can also be helpful to understand possible causes for the inability of the models to make highly accurate predictions for these groups.

The presence of iodine atoms is a dominant characteristic of Group 1 molecules as evidenced by the large descriptor importance and positive difference for the number of iodine atoms (\textit{nI}). These molecules are also distinguished having a larger molecular weight (MW) and higher than typical numbers of bromine atoms (\textit{nBr}). Group 2 molecules have a larger number of rotatable bonds (\textit{nRot}) compared to the others. Having large values for the number of bonds (\textit{nBonds}), the number of SP3 hybridized carbons bonded to two other carbons (\textit{C2SP3}), and the number of single bonds (\textit{nBondsS}) is indicative of the long carbon chains in these molecules which we can visually deduce. For Group 3, having higher atomic polarizability (\textit{apol}), high numbers of basic groups (\textit{nBase}) and higher numbers of acidic groups (\textit{nAcid}) are distinguishing characteristics. In combination, these factors provide an indication that these molecules have a tendency to interact with molecules of the same kind and with the solvent molecules in a solution. Therefore, one can argue that the prediction difficulty for Group 3 molecules is likely due to deficiencies in the current molecular representations to account for complex solute-solute and/or solute-solvent interactions resulting due to the large number of acidic and basic groups.


\subsection{\textbf{Measuring DoA Size}}\label{sec:doa-size}
Once we have established the DoA thresholds for a given property prediction model, we can leverage this information to determine the DoA on any molecular dataset for which we want to make predictions. To demonstrate the practical application of this technique, we sampled 10,000 random molecules from PubChem's collection of compounds~\footnote{https://ftp.ncbi.nlm.nih.gov/pubchem/Compound/Extras/CID-SMILES.gz}. We aim to demonstrate how the choice of training data for a property prediction model can have a significant impact on the size of the DoA of the resulting model. The size of the DoA is defined as the number of molecules in a target dataset that lie inside the domain of interest. This will be the number of molecules for which the model can be expected to achieve a target degree of accuracy. A model with a larger DoA is more generalizable and can be applied with confidence to larger number and diversity of molecular structures. For this purpose we leverage three training sets differing in size and diversity: our full training set (20,229 molecules), the Cui dataset (9933 molecules) and the Delaney dataset (1117 molecules).
The complexity differences in these datasets in terms of structural properties are given in Table S2 in the Supporting Information.

\begin{figure*}[!t]
     \centering
         \begin{subfigure}[b]{0.48\textwidth}
         \centering
         \includegraphics[width=\textwidth]{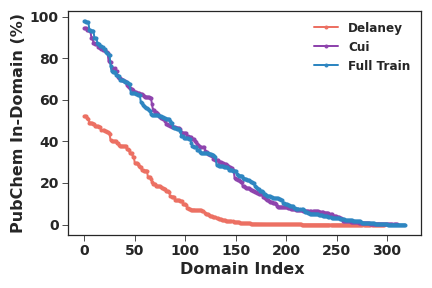}
         \caption{}
         \label{fig:domain-size-a}
     \end{subfigure}
     \begin{subfigure}[b]{0.48\textwidth}
         \centering
         \includegraphics[width=\textwidth]{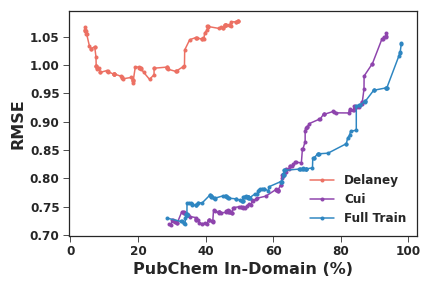}
         \caption{}
         \label{fig:domain-size-c}
     \end{subfigure}

    \caption{
    (a) Percentage of in-domain molecules in the PubChem set based on the DoA determined using models trained on different datasets.
    (b) RMSE of predictions on our test set by models trained on different train sets versus the  percentage of total PubChem molecules which are in-domain for different DoAs.
    }
    \label{fig:domain-size}
\end{figure*}

We would hypothesize that domains of applicability identified using a larger and more diverse training set such as our full training set or the Cui dataset would encompass a larger number of molecules than the domains identified using a smaller dataset.
We can observe this by measuring the number of PubChem molecules that are found to be in-domain for models trained using the three training sets. In Figure \ref{fig:domain-size-a}, we plot the number of in-domain PubChem molecules versus the domain index for the three datasets. As hypothesized, the number of PubChem molecules that fall inside the domains determined by the less diverse Delaney dataset is smaller than the number of molecules inside the domains determined using more diverse datasets.

Each subsequent domain in our DoA analysis is associated with an increasing degree of expected model accuracy. We can quantify the expected degree of accuracy for a given domain by measuring the model performance on a test set of molecules that fall within each domain. We aim to determine the relationship between the expected predictive accuracy on the in-domain data and the size of the in-domain data, as measured by the number of in-domain molecules from the PubChem sample. Using the full train set, the Cui dataset and the Delaney dataset as training sets, we trained three ExtraTreesRegressor models and made predictions for molecules in our test set. 
In Figure \ref{fig:domain-size-c}, we plot the RMSE of our test molecules belonging to a certain domain versus the number of the 10K sample of PubChem molecules that fall in to the same domain. This shows the expected fraction of the external PubChem molecules which fall into different domains and for which we should expect a given degree of accuracy as measured by the RMSE.
For the region where the number of in-domain molecules is the same for all three datasets, in general, molecules selected by the full train set and the Cui domains have significantly higher prediction accuracies than the molecules selected by Delanay's DoA. This shows that models trained on larger more diverse training sets can both be applied across a broader range of molecules and be expected to achieve higher degrees of accuracy.

\subsection{DoA Limitations}

Despite DoA being effective in identifying regions in descriptor space with improved prediction accuracies, we find that DoA analysis by itself cannot be used to detect all the outliers in the data. For example, removing 10\% of the data using our outlier-detection method results in a $R^2$ of about 0.90 for the remaining inliers (Figure \ref{fig:ol-detect}). However, when we remove the same amount of test set data based on the DoA approach, we achieve only a cross-validated $R^2$ of 0.82 (see Table \ref{table:doa-table}). This observation indicates that DoA analysis is not able to identify and remove all of the outlier molecules. In the next section we provide more insight into the types of outliers which cannot be identified using structural filters alone.

\subsection{\textbf{Structural and Data Outliers}}\label{sec:doa-select}

\begin{figure*}[!t]
     \centering
     \includegraphics[width=.9\textwidth]{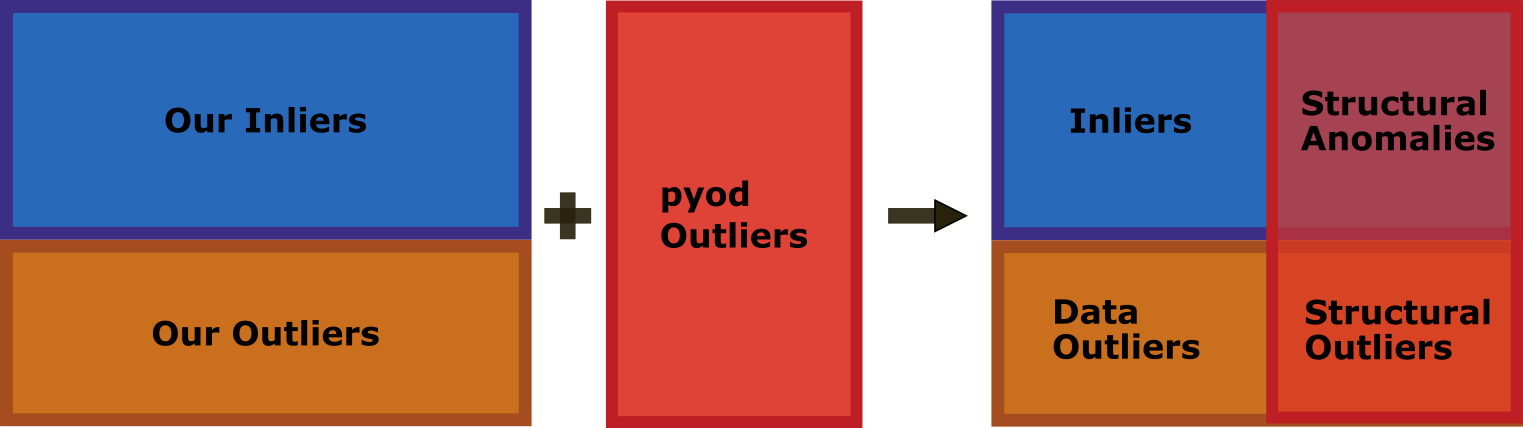}
    \caption{Composition of a molecular dataset in terms of different types of outliers.}
        \label{fig:oltypes}
\end{figure*}

We assume that outliers are of mainly two types: \textit{structural outliers} and \textit{data outliers}. Structural outliers are molecules for which prediction is challenging due to being structurally unusual relative to the training data. Data outliers on the other hand are molecules which are not structurally unusual but still appear challenging for the solubility prediction models. This challenge may occur because of measurement noise in the solubility values, error in the solubility annotations, or difficulties distinguishing the effect of small structural changes on solubility. For these molecules, the target property apparently cannot be predicted accurately by applying the same rules that are valid for most of the other molecules. DoA analysis enables us to identify the molecular features driving the prediction challenges of structural outliers but cannot identify the data outliers because it only takes into account the molecular structure and not the measured property value. There is also a possibility that some molecules may be structurally unusual but are still able to be predicted accurately by the model as some types of structural features may prove less challenging than others. This fact introduces a third class of outliers - \textit{structural anomalies}. In Figure \ref{fig:oltypes} we depict how a dataset can consist of different types of outliers.

To identify structural outliers, we employed 18 relatively fast unsupervised anomaly detection methods implemented in the \textit{pyod} package~\footnote{https://pyod.readthedocs.io/}. Refer to the supporting information for more details on the anomaly detection process. The molecules that get classified as outliers by two or more of the anomaly detection methods were considered as structural outliers.  We found 2693 structural outliers which were detected by both the pyod method and by our method as outliers. As the number of outliers detected by our method is 4046, the remaining 1353 detected outliers from our method were then considered as data outliers since they do not appear to be structurally unusual but still behave as outliers. The 5765 which are flagged by \textit{pyod} but not detected as outliers by our method are considered to be the structural anomalies.

\begin{figure*}[!t]
     \centering
     \begin{subfigure}[b]{0.49\textwidth}
         \centering
         \includegraphics[width=\textwidth]{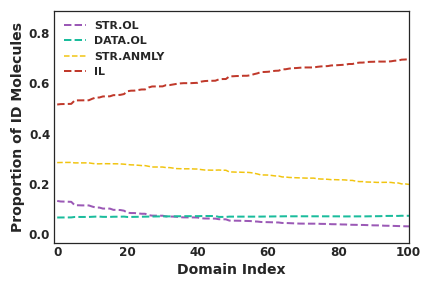}
         \caption{}
     \end{subfigure}
     \begin{subfigure}[b]{0.49\textwidth}
         \centering
         \includegraphics[width=\textwidth]{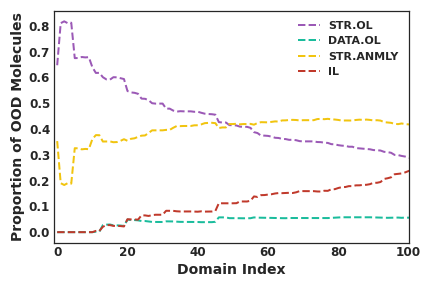}
         \caption{}
     \end{subfigure}
        \caption{Proportion of molecules which are data outliers (DATA.OL), structural outliers (STR.OL), structural anomalies (STR.ANMLY), and inliers (IL) (a) inside and (b) outside the domain of applicability, for first 100 domains. 
        }
        \label{fig:olp-vs-doa-id}
        
\end{figure*}

We next seek to understand the relationships between the different types of the outliers and the molecules that are detected as in-domain and out-of-domain using the DoA method. In Figure \ref{fig:olp-vs-doa-id}, we show the proportion of in-domain and out-of-domain molecules which fall into each of the outlier types for each domain. From this figure, we can see that in-domain molecules consist almost entirely of inliers and structural anomalies indicating that our method to define the domain boundaries is effective at separating well-predicted from poorly-predicted molecules.  

Similarly, we find that the OOD molecules consist primarily of the structural outliers for the highest ranked domains but do start to incorporate some well-predicted structural anomalies as well as a small number of inliers for later domains. This trend indicates the effectiveness of the domain rank-ordering method. From Figure \ref{fig:olp-vs-doa-id} we see that the ability of the threshold values to separate outliers and inliers has reduced by a small amount for domain indices greater than 60. However, as we observed in Figure \ref{fig:fbf}, using only the first 20 domains, we can get rid of highly influential out-of-domain molecules to effect a significant increase in the prediction accuracy. We also demonstrate that the OOD molecules fail to incorporate the data outliers, which are poorly-predicted but not structurally distinct. 

\begin{figure*}[!t]
    \centering
        \includegraphics[width=1\textwidth]{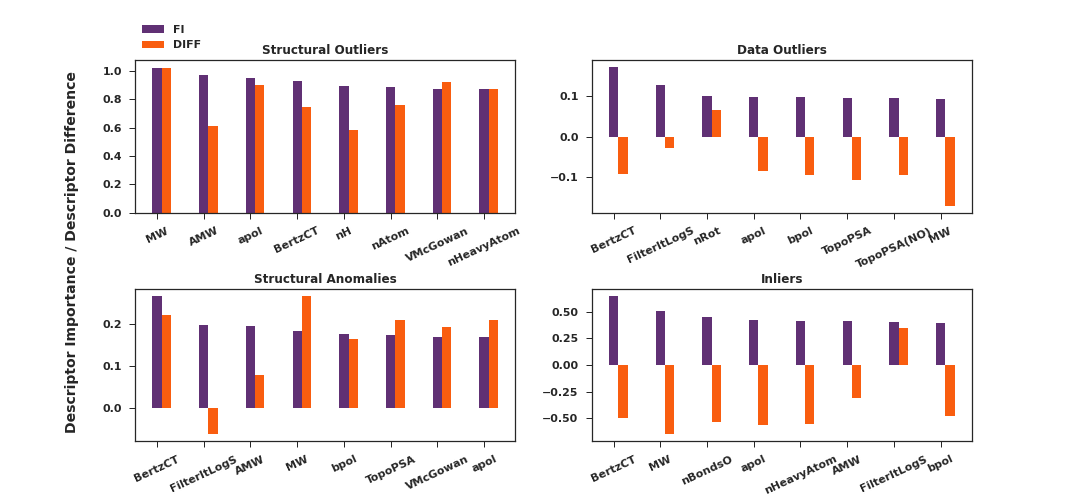}
        \caption{
        The distinguishing characteristics of each type of outliers relative to all other molecules. The purple bars show the descriptor importance scores from the ExtraTreesClassifier when classifying the molecules belonging to each outlier category versus the rest of the molecules. The orange bars show the difference between the mean descriptor values of molecules belonging to each outlier category and the rest of the molecules}
        \label{fig:feat_diff-oltypes}
\end{figure*}

To get a better understanding of the characteristic features of the molecules in different outlier categories, we train a classification model to distinguish each type from the rest of the molecules. We plot the descriptor importance scores and mean feature differences in Figure \ref{fig:feat_diff-oltypes} using the same approach as we did in Section~\ref{sec:subdomains} to analyze the subdomains. We can see that, as expected, structural outliers and structural anomalies can be distinguished by being larger and more complex than the other molecules as well as having higher polarizability. Correspondingly, inlier molecules are found to be significantly smaller and less complex with lower polarizability and higher solubility. For data outliers, the differences in the structural descriptors are smaller than those of the other molecules and appear similar to the inliers. This fact reinforces our hypothesis that data outliers cannot be detected using structural descriptors.

\begin{figure*}[!t]
     \centering
     \begin{subfigure}[b]{0.49\textwidth}
         \centering
         \includegraphics[width=\textwidth]{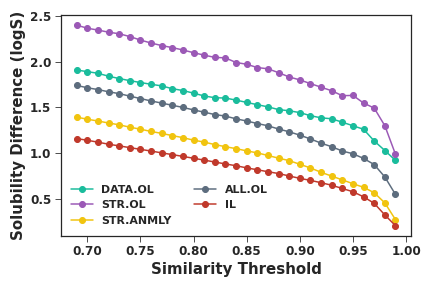}
         \caption{}
     \end{subfigure}
     \begin{subfigure}[b]{0.49\textwidth}
         \centering
         \includegraphics[width=\textwidth]{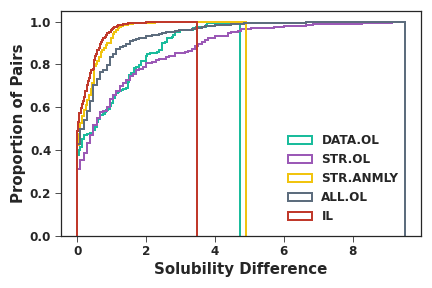}
         \caption{}
     \end{subfigure}
        \caption{ (a) Mean solubility difference versus pairwise cosine similarity for pairs of similar molecules.  (b) Cumulative distribution of the proportion of pairs with cosine similarity $>=$ 0.99 having an absolute solubility difference less than a threshold value. }
        \label{fig:str-data-ols}
\end{figure*}

Based on the findings of our previous work \cite{Panapitiya2022}, we hypothesize that one reason for the existence of the data outliers is limitations in the representation of isomer structures. In some cases even though the molecules are structurally very similar, their solubilities can have significant differences. In Figure \ref{fig:str-data-ols}(a) we plot the mean solubility difference between pairs of molecules, whose molecular descriptor cosine similarity is greater than a threshold value.  In general, we see that the solubility differences between similar molecules follow a fixed ranking of Structural Outliers $>$ Data Outliers $>$ Structural Anomalies $>$ Inliers even as the similarity values approach 1 where the molecules become nearly identical. In Figure \ref{fig:str-data-ols}(b), we focus on the solubility differences of very similar pairs of molecules (cosine similarity > 0.99) and show the distribution of solubility differences for such pairs in each group. The number of pairs with close-to-zero solubility difference is largest for inliers and structural anomalies while structural outliers and data outliers have many similar pairs of molecules with significant solubility differences. 

These results provide an explanation for the relative difficulties of making predictions for the different groups of molecules. For inliers, very similar molecules also tend to have very similar solubilities making it easy for the models to correctly predict their solubilities. A similar patterns holds for the structural anomalies. Even though they are structurally different from the majority, their structurally similar counterparts are likely to have similar solubilities, making it easy for the predictive models to extrapolate to the new structures. Finally, despite not being structurally distinctive compared with the rest of the molecules, the data outliers have significantly larger solubility differences with their structural neighbors than observed for the inliers, explaining the challenge of inferring their solubility. The ability to infer large solubility differences arising from fine-grained structural changes is a limitation of current data-driven predictive ML approaches. It is also possible that some of the large solubility differences observed between similar structures are the result of measurement noise or errors, which would also drive the observed prediction errors for these molecules.

\subsection{Non-linear detection of outliers}

\begin{figure*}[!t]
    \centering
         \includegraphics[width=.8\textwidth]{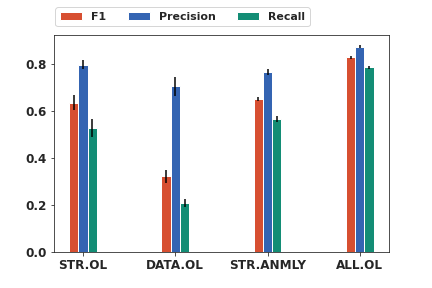}
    \caption{Classification scores corresponding to classifying different types of outliers versus the rest of the molecules using ExtraTreesClassifier models.}
        \label{fig:oltype-classification}
\end{figure*}

One limitation of our DoA approach is that the domain boundaries are each defined using a single descriptor resulting in each domain being defined by a hyper-rectangle in feature space. However, in practice, regions of molecular space are likely described better by non-linear boundaries which are functions of multiple descriptors. We can use machine learning algorithms to understand how such non-linear detection methods might improve on the linear method at the cost of reduced interpretability. In Figure \ref{fig:oltype-classification} we show the classification accuracies of ExtraTreesClassifier models that attempt to distinguish each type of outliers from the rest of the molecules.

The classification results shown in Figure \ref{fig:oltype-classification} indicate that structural outliers and anomalies are both structurally distinguishable from other molecules. Meanwhile, the data outliers are not able to be distinguished from other molecules using structural features alone, which validates our understanding of this type of outlier being related to a combination of structural and property information. Based on the predictability of the structural outliers, we can use the predicted outlierness as a measure to filter in-domain from out-of-domain molecules in a non-linear fashion. However, this approach has a downside of a lack of interpretability relative to our feature-based filtering approach.

\begin{figure*}[!t]
    \centering
         \includegraphics[width=1\textwidth]{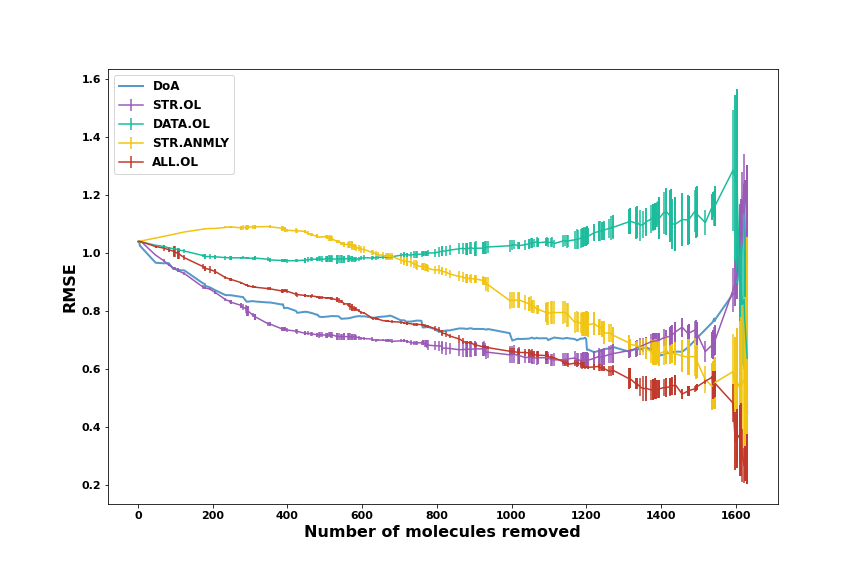}
    \caption{The improvement in solubility prediction RMSE due to the filtering certain numbers of molecules based on the linear DoA method and based on outlier class probabilities from the outlier classifier models. The error bars show the standard deviation of the RMSE values across five randomly initialized models. }
        \label{fig:ol-clsfy}
\end{figure*}

The classifiers were trained to detect inliers and different types of outliers using the same train set which was used to find the DoA thresholds. The trained classifiers can be used to obtain the probabilities for a given molecule to get classified as an inlier or an outlier. The predicted probability of being an outlier is used as metric to separate in-domain (low probability of being an outlier) and out-of-domain (high probability of being an outlier) molecules. If the non-linear patterns detected by the classification model are more effective, we should see that this method can identify applicability domains of the same size but with higher accuracy than the linear DoA method. To test this, we remove the same percentage of molecules corresponding the domain sizes from the linear DoA method and then find the prediction accuracy for the remaining in-domain molecules. Our results are shown in Figure \ref{fig:ol-clsfy} as function of the number of molecules removed.


As evidenced by the decreasing trend in the RMSE, the classifiers trained using structural outliers and the combined set of outliers are effective in identifying molecules which are more difficult to predict.
Interestingly, we find that the linear DoA method performs similarly to the classifiers when removing small numbers of molecules. This provides an indication of the effectiveness of the feature selection of our DoA method at identifying the top set of relevant features for boundary definition. However, when larger numbers of molecules are filtered out to identify the very high-accuracy domains, the benefit of the non-linear filtering becomes clear as the structural outlier classifier begins to outperform the original method. The relative performance of the classification models on the different types of outliers shows that predicting structural outliers is more effective than predicting data outliers or structural anomalies. The failure of the data outlier classifier likely derives from the poor ability of the classifier to identify data outliers based on structural features alone. Meanwhile, the poor performance of the structural anomaly classifier confirms our understanding that structural outliers are structurally unusual but not difficult for the solubility prediction model.

\section{Conclusion}\label{sec13}

We presented a new method to find domains of applicability for which predictive models can be expected to achieve increasing degrees of accuracy. Our method specifies the boundaries of each domain using descriptors or features of the dataset. The effectiveness of our method was shown by calculating the prediction accuracies of the in-domain molecules corresponding to each domain. The accuracies were found to improve as the domains get more restricted. We also proposed methods to get more insight on different domains, which is important for identifying difficult-to-predict molecules and thereby improve model accuracies. We demonstrate that training models based on larger, more diverse datasets leads to increased size of the applicability domains in structural feature space and to improved generalizability of the models to new datasets. Finally, we explore the relationship between structurally-defined applicability domains and molecules which behave as predictive outliers and identify three classes of such outliers. These results indicate that structural information alone is not sufficient to identify the model failure modes.  We demonstrate that training classifiers using the outlier groups is also an effective way to detect applicability domains and that the non-linear thresholding of such models can improve on the linear DoA determination method at the cost of reduced interpretability of the domain boundaries.

\section*{Supplementary information}

Data preparation details, KMeans clsutering, Unsupervised Anomaly Detection, Complexities of PNNL, Cui and Delaney datasets, Sample molecules from subdomains.

\section*{Acknowledgments}
This work was supported by Energy Storage Materials Initiative (ESMI), which is a Laboratory Directed Research and Development Project at Pacific Northwest National Laboratory (PNNL). PNNL is a multiprogram national laboratory operated for the U.S. Department of Energy (DOE) by Battelle Memorial Institute under Contract no. DE- AC05-76RL01830.

\bibliographystyle{unsrtnat}
\bibliography{references}
\end{document}